
\input phyzzx
\def\bau{{\it BAU}}
\def\cp{{\it CP}}

\def\b{{\it B}}

\def\sm{ standard model}
\def\ew{ electroweak phase transition}

\def\su21{$SU(2) \times U(1)$}
\def\su2{$SU(2) $}

\def\del{\Delta(\omega)}

 %
 %
\def\prdabbr{{\it Phys.\ Rev.}\  D}
\def\plbabbr{{\it Phys.\ Lett.}\ B}
\def\npbabbr{{\it Nucl.\ Phys.}\ B}

\def\rmpabbr{{\it Rev.\ Mod.\ Phys.\ }}
\def\mplaabbr{{\it Mod.\ Phys.\ Lett.} \ A }
\def\prd#1#2#3{\prdabbr\ #1\ (19#2)\ #3}
\def\plb#1#2#3{\plbabbr\ #1\ (19#2)\ #3}
\def\npb#1#2#3{\npbabbr\ #1\ (19#2)\ #3}

\def\rmp#1#2#3{\rmpabbr\ #1\ (19#2)\ #3}
\def\mpla#1#2#3{\mplaabbr\ #1\ (19#2)\ #3}
\REF\sakharov{ A.D.~Sakharov {\it JETP Lett.}{\bf 5}, 24, (1967).}
\REF\lindek {D.A.~Kirzhnits and A.D.~Linde, \plb{72}{72}{471}.}
\REF\gleiser{M.~Gleiser, preprint DART-HEP-94-02 (1994) and
references therein.}
\REF\krs {V.A.~Kuzmin, V.A.~Rubakov and M.E.~Shaposhnikov,
\plb{155}{85}{36}.}
\REF\larryvel {L.~McLerran, B.-H.~Liu and N.~Turok,
\prd{46}{92}{2668}.}
\REF\welklim{V.V.~Klimov, {\it Sov.\ Phys.\ JETP} 55 (1982)
199. H.A.~Weldon, \prd{26}{82}{1394}.}
\REF\phase {M.~Dine, P.~Huet, R. G.~Leigh, A.~Linde and D.~
Linde, \plb{283}{92}{319}; \prd{46}{92}{550}.}
\REF\carrington {M.~Carrington, \prd{48}{93}{3836}.}
\REF\stablewall {P.~Huet, K.~Kajantie, R.G.~Leigh, L. McLerran and
B.-H. Liu, \prd{48}{92}{2477}.}
\REF\pat{ P.~Huet, Proceedings of the 1992 Fermilab Meeting of the
DPF Division of the APS, C.H.~Albright, P.H.~Kasper, R.~ Raja and
J.~Yoh, eds., World Scientific, Singapore, 1993, p. 1401.}
\REF\klink{N.~Manton, \prd{28}{83}{2019}; F.~Klinkhamer and
N.~Manton, \prd{30}{84}{2212}.}
\REF\bound {M. E.~Shaposhnikov, \npb{287}{87}{757}.}
\REF\latest{ M.~Dine, P.~Huet and
R.~Singleton,~Jr., \npb{375}{92}{625}.}
\REF\esp{J.R.~Espinosa, M.~Quir\'os and F.~Zwirner,
\plb{314}{93}{206}.}
\REF\gpy{D.J.~Gross, R.D.~Pisarski and L.~Yaffe, \rmp{53}{81}{43}.}
\REF\montecarlo{ See,~for example,~J.~Ambjorn, T.~Askgaard, H.~Porter
and M.E.~ Shaposhnikov,
\plb{244}{90}{479}; \npb{353}{91}{346}.}
\REF\smilga{For a review, see V.V.~Lebedev and A.V.~Smilga,
{\it Annals of Physics}, {\bf 202}, 229 (1990) and references therein.}
\REF\gammaexpress {E.~Braaten and R.D.~Pisarski, \prd{46}{92}{1829}.}
\REF\farrar {G.R.~Farrar and M.E.~Shaposhnikov,
preprint CERN-TH-6734/93 (1993). }
\REF\ghop{M.~B.~Gavela, P.~Hern\'andez, J.~Orloff and O.~P\`ene,
\mpla{9}{94}{795}.}
\REF\us{P. Huet and E. Sather, preprint SLAC-PUB 6479 (1994).}
\pubnum={6492}
\pubtype={T/E/As}
\titlepage
\title{Electroweak Baryogenesis and the Standard Model}
\author{Patrick Huet}
\SLAC
\abstract{Electroweak baryogenesis is addressed within the context of
the standard model of particle physics. Although the minimal standard
model has the means of fulfilling the three Sakharov's conditions, it
falls short to explaining the making of the baryon asymmetry of the
universe. In particular, it is demonstrated that the phase of the CKM
mixing matrix is an insufficient source of {\it CP} violation. The
shortcomings of the standard model could be bypassed by enlarging the
symmetry breaking sector and adding a new source of {\it CP}
violation.}
\vskip 1.5cm
\centerline {\it To appear in Proceedings}
\centerline {\it of the First International Conference}
\centerline {\it on Phenomenology of Unification from}
\centerline {\it Present to Future, Roma, ITALY -- March 23-26,1994}
\endpage
\noindent{\fourteenpoint \bf Introduction}

 The origin of the baryon asymmetry of our universe (\bau ) is a
fundamental question of modern physics. A. Sakharov\refmark\sakharov
established on general ground that particle interactions
might account for the production of the \bau\ at an early epoch of
the universe provided that some of these interactions are \b\
violating processes which operated in a {\it C} and \cp\ violating
environment during a period the universe was in non-thermal
equilibrium.

 The state of the art in particle physics is the \sm\ of the theory
of the interactions between quarks and leptons. \cp\ violation has
been observed and might originate from the mass matrix of the quarks.
\b\ violation is believed to have taken place through
non-perturbative processes of the theory of weak interactions in a
high temperature plasma. Non-equilibrium in particle distributions in
the plasma was generated at the electroweak phase transition.
Although the three Sakharov's requirements are potentially fulfilled
by the \sm\ the latter comes short to explaining the making of the
\bau.

 In what follows, I describe how each of the three Sakharov's
conditions is addressed by the minimal standard model. I briefly
discuss the obstacles to proving or disproving the making of the
baryon asymmetry at the electroweak phase transition using known
ingredients of particle physics.

\noindent{\fourteenpoint \bf I. Departure from Thermal equilibrium}

\noindent{\fourteenpoint \undertext{The \ew} }

\su2\ gauge symmetry was unbroken in the early
universe.\refmark\lindek To argue so, one notes that the Higgs
field was in contact with a thermal bath containing $ M_W^{\pm}$, $Z$
and $t$-quarks whose masses are important parameters of their
equilibrium distributions. On one hand, the Higgs self-interaction
energy, $V(\phi)$, was minimized for an Higgs expectation value $
\phi_m$ of order $v\simeq250$ GeV. On the other hand, the free energy
of the plasma was minimized for $\phi_m = 0$, i.e., in the limit of
massless particles. General arguments of thermodynamics imply that
the Higgs expectation value was then lying at an intermediate value,
at the minimum of the sum of the effective potential and the free
energy of the plasma, $ V(\phi,T)=V(\phi)+ F(M_W,M_t;\phi,T)$.

Calculation of the free energy is a central problem. It is usually
attempted perturbatively, that is, a guess is made as to what a good
approximation of the plasma might be and small corrections
are subsequently added. A popular starting point is {\it a gas of
free particles}; its free energy is
 $$
 F^{\rm free}(M_W,M_t;\phi,T)
 =\sum_{species}\int_0^{M^2}dM'M' \int
 {d^3{\bf k}\over (2\pi)^3}{1\over E}{1\over e^{E/T}-1}
 \eqn\pressure
 $$
 $$
 \qquad\qquad\qquad\simeq  -{103\pi^2\over90}T^4 +{T^2\phi^2
 \over 24}\,\biggl({9 M_W^2+6 M_t^2\over v^2}\biggr) -
 {T \phi^3 \over 12 \pi}\,
 \biggl({9M_W^3 \over v^3}\biggr)+ \ldots \, .
 \eqn\free
 $$
At high temperature, the first two terms in eq. \free\ dominate any
other term in $V(\phi,T)$, hence, $V(\phi,T)$ is minimal for zero
Higgs expectation value, $\phi_m=0$, and {\it electroweak symmetry is
restored at high temperature}. At low temperature, $F^{free}$ is
negligible in respect to $V(\phi,T)$, in which case {\it electroweak
symmetry is broken}. Inevitably, as the universe cooled down to an
intermediate temperature of order $T_c\sim (M_H/M_W)v$ ($\sim 100$
GeV), a phase transition occurred during which the Higgs field
developed a non-zero expectation value which interpolates between $0$
and $v$. The actual mechanism of transition is of crucial importance
for establishing the time scale of non-equilibrium; its determination
requires a complete knowledge of $F(\phi,T)$ beyond the free gas
approximation. There are two ideal possibilities.\foot{For attempts
to interpolate between these two possibilities, see Ref. \gleiser.}

\noindent{\fourteenpoint \undertext{A second order phase transition} }

If the Higgs expectation value evolves continuously from $0$ to $250$
GeV as the universe cools down from $T_c$ to below, the transition is
said to be second order. The time dependence of the Higgs expectation
value induces a departure from equilibrium whose typical time scale
$\tau_{\rm nonequi}$ is of the order of the inverse of the rate of
expansion of the universe, $\tau_{\rm nonequi} \sim H^{-1}\sim
{M_{\rm Planck} \over T^2} \sim 10^{17}/T $. This is about $17$ order
of magnitude slower than a typical microscopic process in the plasma.
The ``baryon per photon" ratio  produced cannot possibly be larger
than $ n_B/s \sim \tau_{\rm equi} / \tau_{\rm nonequi} \sim ({1\over
T})/({10^{17}\over T}) \sim 10^{-17}$, an amount which is too small
to make a significant contribution to the asymmetry observed today,
$(n_B/s)_{Obs.} \sim 10^{-10}$.

\noindent{\fourteenpoint \undertext{A first order phase transition} }

A first order phase transition is one in which the expectation value
of the Higgs develops an instability. A thermal fluctuation triggers
a local transition of $\phi$ from $0$ to a non-zero value $\sim
\phi_m$; this is the phenomenon of bubble nucleation. Some of these
bubbles expand, their interface sweeps the plasma requiring a given
species to becomes massive and to rapidly relax to a new thermal
distribution. This produces a temporary non-equilibrium situation
with a time scale $\tau_{\rm nonequi}$ of the order of $\tau_{\rm
nonequi} \geq {\rm thickness/velocity} \sim 10^{(1-3)}/T $. As this
time scale is comparable to the microscopic plasma scale,
the production of a ``baryon per photon" ratio is allowed in the
neighborhood of the moving bubble wall, up to an amount $n_B/s \sim
10^{-(1-3)}$, far sufficient for the purpose of baryogenesis.

{\it A first order phase transition allows electroweak baryogenesis
to meet the first of Sakharov's criteria}. The uncovering of this
possibility resulted in an extensive study of the dynamics of
expansion of an electroweak bubble.\refmark{\larryvel,\phase} The
main properties of this dynamics reflect heavily the non-equilibrium
structure of the plasma at the bubble interface, and are now fairly
well understood. Bubbles grow to a macroscopic size of order
$10^{12}/T$ until they fill up the universe. This size is far larger
than the microscopic scale of baryogenesis, $\sim 1/\alpha_W T$, and
complications due to the curvature of the wall can be ignored. The
thickness of the interface is of the order $(v M_H^2/M_W^3) 1/T \sim
(10 - 100)/T$, while the terminal velocity of expansion $v_W$ has
been evaluated to be in a non-relativistic range and no smaller than
$\sim 0.02$.\foot{More exactly, in this limit $v_W$ is proportional
to the amount supercooling in the plasma and is sensitive to the
Higgs mass ($\sim 1/ M_H^{3/2}$)and the top quark mass. The value
quoted is for $M_H\sim 60$ GeV and $M_t\sim 180$ GeV.\refmark\phase}
This lower value is the result of a saturation of the wall damping
which is attained when the wall thickness is smaller than the
plasma mean free paths of the relevant gauge bosons and $t$-quarks.
Calculations in this limit are very
reliable.\refmark{\larryvel,\phase} Furthermore, in this range of
velocities, the growth has been shown to be
stable\refmark{\stablewall}. The above considerations lead to a
fairly simple and ``user-friendly" picture of the electroweak phase
transition.

\noindent{\fourteenpoint \undertext{Discussion} }

The current status of our understanding of the electroweak phase
transition leads to the wide-spread belief that it was a first order
transition provided that the Higgs mass not be too
large.\refmark{\phase,\esp} However, unresolved calculational
difficulties due to the non-abelian structure of the gauge sector
prevents to establish reliably the order of the transition. The
essence of these difficulties can be formulated in the following
way.\refmark\pat

Computation of the free energy of a nearly massless plasma beyond the
approximation of a gas of free particles \pressure\ is needed.
Difficulties arise because of large energy corrections from mutual
particle interactions. More specifically, gauge bosons $W$'s and
$Z$'s follow the Bose-Einstein distribution $(\exp E/T -1)^{-1}$. The
plasma contains a large fraction of small momenta ($|{\vec
k}|< M$) gauge bosons, which diverges as $\int d^3{\vec k}\,\,T/M$ in
the massless limit. Multiple gauge interactions between these quanta
contribute to the free energy an amount $(T/M)g_W^2(T/M)g_W^2 \ldots$.
The sum of these contributions, denoted $\Delta F$, can be expressed
as a series expansion in powers of $g_W^2T/M$
 $$\Delta F  = \sum_n\biggl( {g^2_W T\over M}\biggr)^n f_n(M/T) \, .
         \eqn\corrections
 $$
These corrections are expected to affect the quality of the order of
the transition as it has been understood that a first order
transition occurs only as the result of the presence of this large
number of low momenta gauge bosons in the plasma :\foot{More
specifically, the term responsible for the first order structure is
the cubic term in \free.} {\it a reduction in their number implies a
phase transition more weakly first order}.

A leading approximation to the free energy which better accounts for
the contribution of the longitudinal components of the gauge bosons
is the free energy of a {\it gas of quasiparticles}. These
quasiparticles, or ``plasmons",\refmark\welklim are collective
excitations in the plasma and arise from the ``dressing" of the low
momenta ($|{\vec k}|< M$) longitudinal modes with large momenta
($|{\vec k}| \sim T $) modes. The quasiparticle approximation is
obtained by simply substituting the thermal distribution of a free
boson with the one of a free quasiparticle
 $$ {1\over e^{\sqrt{{\vec k}^2+M^2}}-1}\qquad\longrightarrow\qquad
 {1\over e^{\sqrt{{\vec  k}^2+M^2+M_D^2}}-1} \, ,
 \eqn\cutoff
 $$
$M_D$ is the Debye mass $M_D\simeq g_W T$. According to \cutoff, the
fraction of the population of longitudinal (quasi-)gauge bosons with
momentum $|{\vec k}|< M << T$ is reduced in respect to the
corresponding free gauge boson population, as $T/M \to
T/(M^2+M_D^2)^{1/2}\sim 1/g_W$. It is finite ( no infrared
divergence!) and sufficiently small to convert the infrared expansion
\corrections\ of the corresponding interaction energy $\Delta F$,
into a computable series expansion in powers of $g_W$. Making the
substitution \cutoff\ in expression \pressure\ to obtain the leading
contribution $F_{quasi}^{free}$ from $F^{free}$,\refmark\pat leads to a
corresponding suppression of the contribution of the longitudinal
modes to the strength of the first order transition.\foot{ The cubic
term in \free\ becomes ${9 T \over 12 \pi} ( (\phi{ M_W \over v})^2+
M_D^2)^{3/2}\sim {9 T M_D^3\over 12 \pi}+ {\cal O} (\phi^2/T^2)$, for
$\phi \leq \phi_m$.} Resummation methods have been successful to
account for these effects.\refmark{\carrington,\phase}

Interactions among the transverse components of the gauge fields and
between them and other gauge fluctuations in the plasma are
believed\refmark\gpy to generate an effective ``magnetic mass", $M_M
=\# g_W^2T$, which cuts off the distribution of low momenta modes in
a way similar to substitution \cutoff. There is no known method of
computing this mass reliably. Furthermore, contributions to $\Delta
F$ involving transverse gauge bosons cannot be accounted for with
the ``quasiparticle" method: making the analog substitution \cutoff\
in \corrections\ leads to a finite but {\it non-perturbative} series
unless $\#$ is shown to be sufficiently large, in which case,
expression \free\ yields a second order or a very weakly first order
transition. This sort of analyses have been performed using
resummation techniques and are fairly inconclusive. Expansion about a
gas of free quasi-particles is not adequate. Other methods ought to
be developed. Currently, no technique has turned to be successful to
strictly establish the order of the electroweak phase transition.

\noindent{\fourteenpoint \bf II. Baryon number violating processes}

An ideal setting for the baryon number violating processes is as
follows. (a)$B$-violation occurs at an infinite rate
($\Gamma_B^u=\infty$) in the unbroken phase in order to produce an
asymmetry in the region of the plasma disturbed by the moving
interface of a bubble of broken phase. (b)$B$-violation is turned off
in the broken phase ($\Gamma_B^u=0$) in order to preserve inside the
bubble, the \bau\ produced. In practice, this optimal situation is
not realized.

\noindent{\fourteenpoint \undertext{Anomalous baryon number violation }}

The \b\ violating processes of the standard model result from the
conjunction of the non-trivial topological structure of an \su2\
gauge theory and of the chiral \su2\ anomaly. The former can be
formulated in expressing the existence of an infinite number of
distinct sectors in the configuration space which are mapped to each
other by large gauge transformations. These distinct sectors are
characterized by an integer $N_{CS}$, the Chern-Simons number. A
physical process interpolates between two such sectors provided it
has a space-time integrated value of $F{\tilde F}(x)$ proportional to
the difference between the Chern-Simons numbers of the initial and
final sectors $\int dx^4 F{\tilde F}(x) =1 /32\pi^2\,\, \Delta N_{CS}$.
These processes conspire with the anomalous baryon and lepton
currents ( $ \partial j_\mu^{B,L}(x)=(3/ 32\pi^2)F{\tilde F}(x) $
), to produce a net change of the baryon and lepton numbers $\Delta B
= \Delta L = 3 \Delta N_{CS}$. The above formal arguments do not
guarantee the existence of these anomalous processes. However, such
processes are believed to have taken place in the high temperature
plasma of the early universe.
\hfil\eject
\noindent{\fourteenpoint \undertext{Finite temperature baryon number
violation} }

In the broken phase,\refmark\klink adjacent topological sectors are
separated by an energy barrier at the top of which sits the sphaleron
configuration whose energy is proportional to the expectation value
of the Higgs condensate, $E_{sph}\sim g_W \phi_m/\alpha_W$ and whose
typical size is of order $1/g_W \phi_m$. At zero temperature, the
barrier is too high $E_{sph} \simeq 10$ GeV and only instantons can
achieve a transition at a negligible rate. At high temperature,
relevant gauge-Higgs modes are thermally populated and sphalerons
have a non-zero probability of being produced. Baryon number
violation occurs but is Boltzmann suppressed,\refmark\krs $\Gamma_B^b
\propto T^4 e^{-E_{sph}/T}$.

In the unbroken phase, computation of the rate is a difficult task
because of the difficulty of taking into account interactions between
the very large low momenta modes of the massless gauge bosons, as
briefly discussed in the previous section. However, general
principles provide a fair understanding.\refmark\gpy There is a
natural length scale $\xi$ in the plasma, expected to be of order
$1/g_W^2T$. On dimensional ground, a non-trivial configuration of
spatial size $\rho$, has energy $E \sim 1/\alpha_W \rho$. The lowest
energy configuration has the smallest spatial extension which cannot
be smaller than $\xi$. Hence, the least energy configuration is
expected to have energy $1/\alpha_W \xi \sim T$. We conclude that
transitions between the different sectors {\it might} occur at an
unsuppressed rate, that is $\Gamma_B^u=\kappa (\alpha_W T)^4$. No
reliable analytic method of computation currently exists. Numerical
simulations have been performed.\refmark\montecarlo Such studies are
difficult and currently suggest the range $\kappa = 0.1-1$.

\noindent{\fourteenpoint \undertext{Discussion} }

Let us compare the actual situation with the ideal one we promoted
earlier: $\Gamma^u_B=\infty$, $\Gamma^b_B=0$.

In the unbroken phase, $B+L$ violating processes\foot{These processes
are $B-L$ conserving.} occur at an unsuppressed rate $\Gamma_B^u \sim
(\alpha_W T)^4$. During a long period prior the phase transition, as the
temperature decreased from about $10^{13}$ GeV to $100$ GeV, these
processes were in thermal equilibrium, $\Gamma_B^b \gg H T^3$, and
capable to wipe-out any pre-existing $B+L$ asymmetry.\refmark\krs\
This is a major constraint on  models of early baryogenesis and has
been the main motivation for contemplating electroweak baryogenesis.

In the broken phase, $\Gamma^b_B\propto T^4 e^{-E_{sph}/T}$. Not to
loose the \bau\ produced, requires to tune this rate to  no more than
one baryon violating process per unit volume in a lifetime of the
universe, $\Gamma_B^b \ll H T^3$. The structure of $\Gamma_B^b$ and
of $E_{sph} \sim g_W\phi_m/\alpha_W$ transform this condition into a
restriction on the magnitude of the Higgs expectation value: $ \phi_m
\geq T$. In the minimal standard model, $ \phi_m$ is inversely
proportional to the square of the Higgs mass so that the above
condition translates into an upper bound on the Higgs
mass.\refmark\bound The latest estimates of this
bound\refmark{\latest,\phase} yield $M_H < 45 $ GeV, a value ruled
out experimentally. This bound is a direct challenge to implementing
electroweak baryogenesis in the minimal standard model. However, in
almost any extension of the standard model, the relation between the
Higgs expectation value and the Higgs mass involves additional
parameters, which, in some cases,\refmark\esp may result in a less
constraining bound compatible with current experimental data.

\noindent{\fourteenpoint \bf III. \cp\ violation}

In this last section, I confront the third of Sakharov's conditions to
the standard model. More specifically, I establish the impossibility
of implementing the complex phase of the CKM matrix as the source of
\cp\ violation for baryogenesis.  This discussion follows closely a
recent work done in collaboration with E. Sather.\refmark\us

\noindent{\fourteenpoint \undertext{A possible mechanism} }

The standard model possesses a natural source of \cp\ violation
contained in the phase of the CKM matrix. Whether the latter
participated to the making of the \bau\ is a fundamental question
which was addressed for the first time in a proper physical context
by Farrar and Shaposhnikov.\refmark\farrar These authors proposed a
simple mechanism of baryogenesis based on the observation that as the
wall sweeps through the plasma, it encounters equal numbers of quarks
and antiquarks which reflect asymmetrically as a result of the
presence of $CP$-violation. This mechanism leads to an excess of
baryons inside the bubble and an equal excess of antibaryons outside
the bubble. Ideally, the excess of baryons outside is eliminated by
baryon violating processes while the excess inside is left intact,
leading to a net {\it BAU}. Assuming ideal conditions, an upper bound
for the ``baryon-per-photon" ratio can be
derived:\refmark{\farrar,\us} $n_B/s\leq 10^{-2}\times \alpha_W \times
v_W \times \langle\del\rangle_T\ $. The whole calculation of the
baryon asymmetry now reduces to the determination of a suitable
thermal average of the left-right reflection asymmetry $\del= {\rm
Tr}\bigl( |R_{LR}|^2-|{\bar R}_{RL}|^2\bigr)$, that is, the
probability of a L-handed quark reflecting as a R-handed quark minus
the probability for the \cp\ conjugate process, summed over all
quarks.

The non-trivial structure of the phase space is contained in the
velocity factor $v_W$ which reflects the departure from thermal
equilibrium and the factor of $\alpha_W$ which reflects the vanishing
of any asymmetry unless interactions with the $W$ and $Z$ bosons in
the plasma are taken into account in the propagation of the quarks.
In addition, the $CP$-odd quantity $\Delta(\omega)$ vanishes unless
flavor mixing interactions occur in the process of scattering. This
requires to taking into account interactions with the charged $W$ and
Higgs bosons. Furthermore, gluon interactions ought to be included
for they strongly affect the kinematics of the quarks. At first, this
might appear an insurmountable task. However, Farrar and Shaposhnikov
suggested that all the relevant plasma effects can consistently be
taken into account by describing the process as a scattering of
suitably-defined\foot{These quasiparticles are the fermionic
equivalent of the quasiparticles considered in section 1.} {\it
quasiparticles}.\refmark{\welklim,\smilga}

\noindent{\fourteenpoint \undertext{$CP$ violating observable} }

$\del$ is a \cp\ violating observable. It is known that a
$CP$-violating observable is obtained by interfering various
amplitudes with different $CP$ properties. Farrar and Shaposhnikov
proposed to describe the scattering of quasiparticles as completely
quantum mechanical, that is, by solving the Dirac equation in the
presence of a space-dependent mass term. In particular, they
identified the source of the phase separation of baryon number as
resulting from the interference between a path where, say, an
$s$-quark (quasiparticle) is totally reflected by the bubble with a
path where the $s$-quark first passes through a sequence of flavor
mixings before leaving the bubble as an $s$-quark. The $CP$-odd phase
from the CKM mixing matrix encountered along the second path
interferes with the $CP$-even phase from the total reflection along
the first path. Total reflection occurs only in a small range of
energy of width $m_s$ corresponding to the mass gap for strange
quarks in the broken phase. This leads to a phase space suppression
of order $m_s/T$, in which case, the ``baryon-per-photon" ratio
becomes $n_B/s \simeq 10^{-3} \alpha_W (m_s / T) \bar\Delta \simeq
10^{-7}\,\times\, {\bar \Delta}$. This estimate requires ${\bar
\Delta}$, the energy-averaged value of the reflection asymmetry, to
be at least of order $10^{-4}$ in order to account for the baryon
asymmetry of the universe, a value claimed to be attained in Ref.
\farrar.

In Ref. \ghop, it was pointed out that the above analysis ignores the
quasiparticle width, or damping rate, embodied by the imaginary part
of the thermal self-energy $\gamma$. The width  has been computed at
zero momentum as $\gamma \simeq 0.15 g_s^2 T \simeq 20$
GeV.\refmark\gammaexpress These authors made the important
observation that this spread in energy is much larger than the mass
gap $\sim m_s$ in the broken phase, and as a result largely suppresses
the reflection process. A detail of their calculations is to appear
soon.

What follows contains a summary of a recent and alternative
analysis\refmark\us which {\it fully} takes into account all relevant
properties of a (quasi-)quark, as it propagates through the bubble
wall.

\noindent{\fourteenpoint \undertext{Quantum coherence}}

A Dirac equation describes the relativistic evolution of the
fundamental quarks and leptons. Its applicability to a quasiparticle
is reliable for extracting on-shell kinematic information, but one
should be cautious in using it to extract information on its
off-shell properties. A quasiparticle is a convenient bookkeeping
device for keeping track of the dominant properties of the
interactions between a fundamental particle and the plasma. For a
quark, these interactions are dominated by tree-level exchange of
gluons with the plasma. It is clear that these processes affect the
coherence of the wave function of a propagating quark. To illustrate
this point, let us consider two extreme situations.\hfil\break
$\bullet$ {\it The gluon interactions are infinitely fast}. In
this case, the phase of the propagating state is lost from point to
point. A correct description of the time evolution can be made in
terms of a totally incoherent density matrix. In particular, no
interference between different paths is possible because each of them
is physically identified by the plasma. As a result, no
$CP$-violating observable can be generated and
$\Delta(\omega)=0$.\hfil\break
$\bullet$ {\it The gluon interactions are extremely slow}. The
quasiparticle is just the quark itself and is adequately described by
a wavefunction solution of the Dirac equation, which corresponds to a
pure density matrix. In particular, distinct paths cannot be
identified by the plasma, as the latter is decoupled from the
fermion. This situation was implicitly assumed in the FS mechanism.
This assumption, however, is in conflict with the role the plasma
plays in the mechanism, which is to provide a left-right asymmetry as
well as the necessary mixing processes. In addition, this assumption
is in conflict with the use of gluon interactions to describe the
kinematical properties of the incoming (quasi-)quark.

The actual situation is of course in between the two limits above.
The quasiparticle retains a certain coherence while acquiring some of
its properties from the plasma. Whether this coherence is sufficient
for quantum mechanics to play its part in the making of a
$CP$-violating observable at the interface of the bubble is the
subject of the remaining discussion.

The damping rate $\gamma$ characterizes the degree of coherence of
the quasiparticle. It is a measure of the spread in energy, $\Delta E
\sim 2\gamma,$ which results from the ``disturbance'' induced by the
gluon exchanged between the quark and the plasma. From the
energy-time uncertainty relation, $1/(2\gamma)$ is the maximum
duration of a quantum mechanical process before the quasiparticle is
scattered by the plasma.  During this time, the quasiparticle
propagates over a distance $\ell = v_g/ 2\gamma \simeq 1/ 6\gamma
\simeq (100$ GeV$)^{-1}$,
 where $v_g$ is the group velocity of the quasiparticle ( $\sim 1/3$).
The distance $\ell$ was introduced in Ref. \us\ as the {\it coherence
length}. The concept of coherence length leads to a straightforward
description of the decoherence that occurs during the scattering off
a bubble.

\noindent{\fourteenpoint \undertext{Limited coherence and bubble
reflection}}

To understand the impact of the limited coherence of the
quasiparticle on the physics of scattering off the bubble, it is
useful to remember the mechanism of the scattering of light by a
refracting medium. According to the microscopic theory of reflection
of light, the refracting medium can be decomposed into successive
layers of scatterers which diffract the incoming plane wave, the
thickness of a layer reflects the mean interspacing between
scatterers $d$. The first layer scatters the incoming wave as a
diffracting grid. Each successive layer reinforces the intensity of
the diffracted wave and sharpens its momentum  distribution. As more
layers contribute to the interference, the diffracted waves resemble
more and more the full transmitted and reflected waves. This occurs
{\it only} because the wave penetrates the wall {\it coherently} over
a distance large compared to the interspacing of the scattering
sites $\ell \gg d$.

Inspired from the above, we slice the bubble into successive
layers which scatter the incoming wave. The wavefunction for a
quasiparticle reflected from the bubble is the superposition of the
waves reflected from each of the layers. The bulk of the broken phase
can be viewed as a distribution of scatterers whose mean spacing $d$
is the inverse quark mass. However, in contrast with the scattering
of light by a refractive medium, the coherence length $\ell \sim {1 /
100}{\rm GeV}^{-1}$ of the quasiparticle is much shorter than the
interdistance between scatterers $d \sim 1/m_q$. That is, the
quasiparticle does not penetrate the bubble coherently over a
distance large enough to be fully reflected and its reflection
amplitude is suppressed by the ratio $\ell/d \sim m_q/(100$ GeV) $
\ll 1$. In addition, flavor changing processes have to occur along
some reflection paths. Their amplitudes are suppressed by quark masses
and mixing angles; the resulting mean interspacing $d_F$ between
scatterings is then much larger than the coherence length: $\ell \ll
d_F$. These processes are rare events inside the outer layer of the
bubble where coherent reflection takes place and their contribution
to the reflection amplitude is suppressed by $\ell/d_F \ll 1$.

The generation of a \cp\ violating observable results from
interference of reflected waves and necessarily involves several
flavor-changing scatterings inside the bubble in order to pick up the
complex phase of the CKM matrix and several chirality flips.
Altogether, the \cp\ asymmetry $\del$, produced by the scattering
when decoherence is properly taken into account, will be smaller than
the amount found by Farrar and Shaposhnikov by several factors of
$\ell/d$ and $\ell / d_{{\rm F}}$.

{}From these physical considerations, it is easy to elaborate
quantitative methods for computing the reflection of quasiparticles
with a finite coherence length.

A simple model is obtained by expressing that when a quasiparticle
wave reaches a layer a distance $z$ into the bubble, its amplitude
will have effectively decreased by a factor $\exp(-z/2\ell)$. We can
take this into account by replacing the step-function bubble profile
with a truncated profile such as $ m_q(z) =  m_qe^{-z/\ell}$, $z>0$ and
$m_q(z) = 0$, $z<0$. The analog in the theory of light scattering is
the scattering of a light ray by a soap bubble. It is clear that
truncating the bubble in this way renders the bubble interface
transparent to the quasiparticle.

Another method of computing $\Delta(\omega)$ is to solve an effective
Dirac equation in the presence of the bubble, including the
decoherence that results from the imaginary part of the
quasiparticle self-energy. Green's functions are extracted which
allow to construct all possible paths of the quasiparticles
propagating in the bulk of the bubble, each path being damped by a
factor $\exp(-{\cal L}/2\ell)$ where ${\cal L}$ is the length of the
path. Paths occurring within a layer of thickness $\ell$ dominate the
reflection amplitudes, in agreement with the previous considerations.

In Ref. \us, $\Delta(\omega)$ was computed using both methods. They give
results qualitatively and quantitatively in close agreement. They
yield the ``baryon per photon" ratio
 $$
\Biggl|{n_B \over s}\Biggr| <  10^{-26} \ .
 $$
It is $16$ orders of magnitude too small to account for a significant
fraction of the asymmetry observed today, $(n_B/s)_{Obs.} \sim
10^{-10}$.

\noindent{\fourteenpoint \undertext{ Discussion} }

The arguments developed above are powerful enough to establish more
generally that {\it the complex phase allowed in the CKM mixing matrix
cannot be the source of $CP$ violation needed by any mechanism of
electroweak baryogenesis in the minimal standard model or any of its
extensions}. Indeed, the generation of a $CP$-odd observable requires
the quantum interference of amplitudes with different $CP$-odd and
$CP$-even properties and whose coherence persists over a time of at
least $1/m_q$.  On the other hand, QCD interactions restrict the
coherence time to be at most $\ell \sim 1/(g_s^2 T)$, typically three
orders of magnitude too small. Because any $CP$-violating observable
proceed through interference between amplitudes with multiple flavor
mixings and chirality flips, the asymmetry between quarks and
antiquarks appears to be strongly suppressed by many powers of $\ell
m_q$ and mixing angles. This line of argument does not rely on the
details of the mechanism considered and can be applied to rule out
any scenario of electroweak baryogenesis which relies on the phase of
the CKM matrix as the only source of $CP$ violation.

\singlespace
{\bf -Acknowledgements-}
 The account on \cp\ violation follows from a collaboration with
E. Sather and enlightening discussions with M.E. Peskin.
\refout \bye